\documentclass[prd,twocolumn,nopacs,floatfix,amsmath,nofootinbib,amssymb,preprintnumbers]{revtex4-2}
\usepackage{graphicx,color,dcolumn,booktabs,diagbox}
\usepackage{txfonts}
\usepackage{cancel}
\usepackage{slashed}
\usepackage{epstopdf}
\usepackage{multirow}
\usepackage{tikz}
\usetikzlibrary{decorations.pathmorphing}
\usepackage{bm}
\usepackage{braket}
\usepackage{pifont}
\usepackage{adjustbox}
\usepackage{rotating}
\usepackage{verbatim}
\usepackage{appendix}
\usepackage{graphicx,xcolor,dcolumn,booktabs}
\usepackage[colorlinks,
            citecolor=blue,
            anchorcolor=red,
            menucolor=red,
            linkcolor=red,
            filecolor=red,
            runcolor=red,
            urlcolor=blue,
            frenchlinks=red]{hyperref}

\graphicspath{{Figures/}}
\begin{document}
\title{Proposed mixing between $2P$ and $1F$ wave charmonia}
\author{Peng-Yu Sun$^{1,2,4}$}\email{sunpy2025@lzu.edu.cn}
\author{Tian-Le Gao$^{1,2,4}$}\email{gaotl2025@lzu.edu.cn}
\author{Zi-Long Man$^{1,2,4}$}\email{manzl@lzu.edu.cn}
\author{Xiang Liu$^{1,2,3,4}$}
\email{xiangliu@lzu.edu.cn}
\affiliation{
$^1$School of Physical Science and Technology, Lanzhou University, Lanzhou 730000, China\\
$^2$Lanzhou Center for Theoretical Physics, Key Laboratory of Theoretical Physics of Gansu Province, Key Laboratory of Quantum Theory and Applications of MoE, Gansu Provincial Research Center for Basic Disciplines of Quantum Physics, Lanzhou University, Lanzhou 730000, China\\
$^3$MoE Frontiers Science Center for Rare Isotopes, Lanzhou University, Lanzhou 730000, China\\
$^4$Research Center for Hadron and CSR Physics, Lanzhou University and Institute of Modern Physics of CAS, Lanzhou 730000, China}

\date{April 2026}

\begin{abstract}

We investigate $2P$-$1F$ mixing in charmonium, focusing on the close-in-mass $\chi_{c2}(2P)$ and $\chi_{c2}(1F)$ states. The conventional tensor force yields negligible mixing, motivating the inclusion of coupled-channel effects. Our unquenched calculation reveals sizable mixing angles of $7.5^\circ$ and $15.4^\circ$. We predict the corresponding two-photon and two-gluon decay widths as key observables for experimental verification. Additionally, we discuss the production of these two $2P$-$1F$ mixed states of charmonium via $\gamma\gamma$ fusion. 
Current data are insufficient to determine the mixing, highlighting the need for precise future measurements to resolve this aspect of charmonium spectroscopy.

\end{abstract}

\maketitle

\section{Introduction}

In the charmonium system, the $S$-$D$ wave mixing scheme is widely employed to describe vector states. A well-known example is the $2S$-$1D$ mixing model proposed by Rosner to explain the decay properties of $\psi(3686)$ and $\psi(3770)$~\cite{Rosner:2001nm}. This scheme can naturally be extended to higher-mass vector charmonia. For instance, in Refs.~\cite{Man:2025zfu}, the large mixing angles observed for $\psi(4220)$ and $\psi(4380)$ are understood via coupled-channel effects, where a $4S$-$3D$ mixing framework is introduced. Furthermore, the previously unclear $3S$-$2D$ mixing pattern between $\psi(4040)$ and $\psi(4160)$ was clarified in Ref.~\cite{Man:2025vmm}, though further experimental tests are still needed~\cite{Wang:2025dur}.

The $S$-$D$ wave mixing mechanism is not limited to charmonium; it also applies to vector bottomonium~\cite{Li:2021jjt,Bai:2022cfz,Li:2022leg,PhysRevD.109.094045,PhysRevD.109.014039} and light-flavor mesons~\cite{Bai:2025knk}. As an important spectroscopic effect, $S$-$D$ mixing plays a crucial role in explaining several puzzling phenomena in hadron spectroscopy~\cite{Man:2025vmm,Wang:2019mhs,Wang:2022jxj,Wang:2023zxj,Peng:2024xui,Li:2021jjt,Bai:2022cfz,Li:2022leg,PhysRevD.109.094045,PhysRevD.109.014039,Man:2025zfu}.

In this work, we mainly focus on the $2P$-$1F$ mixing, which requires both the experimental discoveries and theoretical analyses of the charmonium $2P$ states. Experimentally, in the open-charm channel, Belle observed the $Z(3930)$ structure in $\gamma\gamma\to D\bar D$ and identified it as the $\chi_{c2}(2P)$ candidate on the basis of its production and angular distributions \cite{Belle:2005rte}. Then, Belle observed a structure, commonly referred to as $X(3915)$, in the photon-photon fusion process $\gamma\gamma\to \omega J/\psi$, obtaining $M=3915\pm 3\pm 2~\mathrm{MeV}$ and $\Gamma=17\pm 10\pm 3~\mathrm{MeV}$ \cite{Belle:2009and}.

The establishment of the $2P$ charmonium multiplet has been a gradual achievement of combined experimental and theoretical efforts, highlighting the crucial role of coupled-channel effects in this mass region~\cite{Bai:2026atm}. In 2009, the Lanzhou group identified $Z(3930)$~\cite{Belle:2005rte} and $X(3915)$~\cite{Belle:2004lle} as the $\chi_{c2}(2P)$ and $\chi_{c0}(2P)$ states, respectively, within their quark model analysis~\cite{Liu:2009fe}. This $J^{PC}=0^{++}$ assignment for $X(3915)$ was subsequently supported by the BaBar Collaboration~\cite{BaBar:2012nxg} and adopted in the 2013 edition of the \textit{Review of Particle Physics} (RPP)~\cite{ParticleDataGroup:2012pjm}.

However, this identification was challenged in 2014~\cite{Guo:2012tv}, which questioned the large observed width for $X(3915) \to J/\psi \omega$, the absence of a $D\bar{D}$ decay mode, and the seemingly small mass splitting between the putative $\chi_{c2}(2P)$ and $\chi_{c0}(2P)$ states. Consequently, $X(3915)$ was removed from the $\chi_{c0}(2P)$ entry in the 2016 RPP~\cite{ParticleDataGroup:2016lqr}.

Theoretical work followed to address these puzzles. The Lanzhou group suggested~\cite{Chen:2012wy} that the $Z(3930)$ structure might contain both $\chi_{c0}(2P)$ and $\chi_{c2}(2P)$ components, potentially explaining the missing $D\bar{D}$ signal for $X(3915)$. Regarding the mass splitting, it was argued that significant mass shifts from coupled-channel effects—analogous to the well-known case of $X(3872)$~\cite{PhysRevD.32.189,Kalashnikova:2005ui}—must be considered for $2P$ states above the open-charm threshold. This perspective was quantitatively supported by an unquenched quark model calculation~\cite{Duan:2020tsx}, which demonstrated how such effects could naturally explain the small $\chi_{c0}(2P)$-$\chi_{c2}(2P)$ mass gap while treating $X(3915)$ as the $\chi_{c0}(2P)$ state. This theoretical conclusion was later confirmed by the LHCb Collaboration through an analysis of $B\to KD\bar{D}$ decays~\cite{LHCb:2020pxc}. Consequently, $X(3915)$ was reinstated as the $\chi_{c0}(2P)$ state in the 2022 edition of the RPP~\cite{ParticleDataGroup:2022pth}.

With the establishment of the $2P$ charmonium multiplet, we are naturally led to an interesting yet relatively unexplored issue—the $2P$-$1F$ mixing scheme in charmonium. Specifically, in the $J^{PC}=2^{++}$ sector, the second radial excitation of the $P$-wave state, $\chi_{c2}(2P)$, and the ground $F$-wave state, $1^3F_2$, are predicted to lie close in mass. Such proximity allows them to mix via intermediate hadronic loops according to former research experience~\cite{PhysRev.176.1709,Akama:1976ni,Renard:1978ha,Kinnunen:1978qm,Tornqvist:1979hx,Tornqvist:1981bs,Ono:1983rd,Tornqvist:1995kr,Kalashnikova:2005ui,Barnes:2007xu,Pennington:2007xr,Zhou:2011sp,Zhou:2013ada,Duan:2020tsx,Duan:2021alw,Eichten:1978tg,Eichten:1979ms,Danilkin:2010cc,Danilkin:2009hr,Li:2009ad,Ortega:2009hj}, even if the direct tensor coupling is weak. Neglecting this $2P$-$1F$ mixing could lead to discrepancies in predicting the properties of $\chi_{c2}(3930)$ and other tensor charmonium candidates. Therefore, a systematic study of the $2P$-$1F$ mixing mechanism is necessary to clarify the fine structure of the charmonium spectrum in this energy region.

Here we clarify our notation. In this work, \(\chi_{c2}(2P)\) denotes the
unmixed \(2\,{}^3P_2\) charmonium basis state before the \(2P\)-\(1F\) mixing is
introduced. It should not be directly identified with the observed
\(\chi_{c2}(3930)\). Instead, in our interpretation, the physical mixed state
\(\chi'_{c2}\), which is dominated by the \(2P\) component, is associated with
\(\chi_{c2}(3930)\). The different notations used above follow the
conventions adopted in the corresponding theoretical or experimental references.

This work focuses on the $2P$-$1F$ mixing problem in charmonium spectroscopy. While the tensor term in the effective potential of traditional quark models can in principle mediate $P$-$F$ mixing, its contribution is typically too weak to generate a sizable mixing angle. This calls for the investigation of alternative mechanisms. Recent progress in understanding unquenched effects highlights that high-mass states cannot be accurately described within the quenched approximation. In particular, the coupled-channel mechanism has proven to be crucial, signaling that hadron spectroscopy is entering a high-precision era. Within this framework, we explore whether coupled-channel effects can induce a significant $2P$-$1F$ mixing. Our calculations yield mixing angles of $\theta_{1}=7.5^{\circ}$ and $\theta_{2}=15.4^{\circ}$.

Based on the mixed-state wave functions obtained from the coupled-channel analysis, we predict the corresponding two-photon and two-gluon decay widths. These distinct decay patterns provide key observables for future experimental tests of the $2P$-$1F$ mixing scenario. At present, however, available experimental data remain insufficient for a precise determination of the mixing angle. Our results highlight the need for higher-precision measurements to resolve the $2P$-$1F$ mixing scheme for charmonium sector.

Furthermore, we explore the potential to produce a higher mixing state via $\gamma\gamma$ fusion. The $2P$-$1F$ mixing results in an enhanced coupling of this state to two photons, relative to states without mixing. Therefore, studying its production could offer crucial insights for future experiments like Belle II.

The paper is structured as follows. Following this introduction, Sec.~\ref{sec2} illustrates the theoretical framework for the $2P$-$1F$ mixing scenario. The corresponding numerical results are presented in Sec.~\ref{sec:numerical}. In Sec. \ref{sec4}, we discuss the production of a higher mixing state in $\gamma\gamma$ fusion process.  
Finally, a concise summary concludes the paper.

\section{Theoretical framework for the $2P$-$1F$ mixing scheme}
\label{sec2}

To determine the bare masses of the relevant charmonia, we employ a potential model calculation, which provides essential input for the subsequent coupled-channel analysis. Various potential models, predominantly derived from the Cornell potential, have been used to describe the hadronic mass spectrum \cite{PhysRevD.32.189,Barnes:2005pb,Godfrey:2014fga,Godfrey:2015dva,Lu:2016mbb,Godfrey:2016nwn}. Among these, the Godfrey--Isgur (GI) model \cite{PhysRevD.32.189} is particularly well-suited, as its precision meets the requirements for contemporary studies of the hadron mass spectrum.

The bare charmonium spectrum is obtained from the Godfrey--Isgur (GI) model,
which is a semi-relativistic quenched potential model. Its Hamiltonian is written
as
\begin{equation}
H
=
\left(\mathbf p^2+m_1^2\right)^{1/2}
+
\left(\mathbf p^2+m_2^2\right)^{1/2}
+
V(\mathbf p,\mathbf r),
\end{equation}
where \(m_1\) and \(m_2\) are the constituent quark masses. For charmonium,
they both correspond to the charm-quark mass. The effective potential
\(V(\mathbf p,\mathbf r)\) contains a short-range one-gluon-exchange part
with Lorentz structure \(\gamma^\mu\otimes\gamma_\mu\), and a long-range
confining part with Lorentz structure \(1\otimes 1\). In the nonrelativistic
limit, the interaction between the \(i\)-th and \(j\)-th constituent can be
decomposed as
\begin{equation}
V_{ij}(\mathbf p,\mathbf r)
=
H_{ij}^{\rm conf}
+
H_{ij}^{\rm so}
+
H_{ij}^{\rm hyp}.
\end{equation}

The spin-independent confining interaction is given by
\begin{equation}
H_{ij}^{\rm conf}
=
\left(
\frac{3}{4}c
+
\frac{3}{4}br
-
\frac{\alpha_s(r)}{r}
\right)
\mathbf F_i\cdot\mathbf F_j ,
\end{equation}
where the first two terms represent the linear confinement potential, while the
last term is the Coulomb-type potential generated by one-gluon exchange.

The color-hyperfine interaction reads
\begin{equation}
\scalebox{0.9}{$\displaystyle
H_{ij}^{\rm hyp}
=
\frac{\alpha_s(r)}{m_i m_j}
\left\{
\frac{8\pi}{3}
\mathbf S_i\cdot\mathbf S_j\,\delta^3(\mathbf r)
+
\frac{1}{r^3}
\left[
\frac{3(\mathbf S_i\cdot\mathbf r)(\mathbf S_j\cdot\mathbf r)}{r^2}
-
\mathbf S_i\cdot\mathbf S_j
\right]
\right\}
\mathbf F_i\cdot\mathbf F_j ,
$}
\end{equation}
which consists of a contact spin-spin term and a tensor term.  Here
\(\mathbf S_i\) and \(\mathbf S_j\) denote the spins of the constituent quark
and antiquark, respectively; in the charmonium case they correspond to the
charm quark and the anti-charm quark. 

The spin-orbit interaction is separated into a color-magnetic contribution and a
Thomas-precession contribution,
\begin{equation}
H_{ij}^{\rm so}
=
H_{ij}^{\rm so(cm)}
+
H_{ij}^{\rm so(tp)} ,
\end{equation}
the color-magnetic part is
\begin{equation}
H_{ij}^{\rm so(cm)}
=
-
\frac{\alpha_s(r)}{r^3}
\left(
\frac{1}{m_i}
+
\frac{1}{m_j}
\right)
\left(
\frac{\mathbf S_i}{m_i}
+
\frac{\mathbf S_j}{m_j}
\right)
\cdot\mathbf L\,
\left(\mathbf F_i\cdot\mathbf F_j \right) ,
\end{equation}
whereas the Thomas-precession term takes the form
\begin{equation}
H_{ij}^{\rm so(tp)}
=
-
\frac{1}{2r}
\frac{\partial H_{ij}^{\rm conf}}{\partial r}
\left(
\frac{\mathbf S_i}{m_i^2}
+
\frac{\mathbf S_j}{m_j^2}
\right)
\cdot\mathbf L .
\end{equation}

In the above expressions,
\(\mathbf L=\mathbf r\times \mathbf p\) is the relative orbital angular
momentum. The color operator \(\mathbf F_i\) is defined as
\begin{equation}
\mathbf F_i=
\begin{cases}
\dfrac{\boldsymbol{\lambda}_i}{2}, & \text{for quarks},\\[6pt]
\dfrac{\boldsymbol{\lambda}_i^c}{2}
=
-\dfrac{\boldsymbol{\lambda}_i^{*}}{2}, & \text{for antiquarks},
\end{cases}
\end{equation}
where \(\boldsymbol{\lambda}_i\) are the Gell-Mann matrices. For a color-singlet
meson, the color matrix element is
\begin{equation}
\langle \mathbf F_i\cdot \mathbf F_j\rangle
=
-\frac{4}{3},
\end{equation}
for comparison, the corresponding color factor in a baryon is
\begin{equation}
\langle \mathbf F_i\cdot \mathbf F_j\rangle
=
-\frac{2}{3},
\end{equation}
since the present work deals with charmonium states, we use the mesonic color
factor \(\langle \mathbf F_i\cdot \mathbf F_j\rangle=-4/3\) in the calculation.

Among these spin-dependent interactions, the tensor part of
\(H_{ij}^{\rm hyp}\) is directly relevant to the \(P\)-\(F\) mixing discussed in
this work. It can be written as
\begin{equation}
H_T
=
\frac{4\alpha_s(r)}{3m_i m_j r^3}
\left(
\frac{3(\mathbf S_i\cdot\mathbf r)(\mathbf S_j\cdot\mathbf r)}{r^2}
-
\mathbf S_i\cdot\mathbf S_j
\right) ,
\label{eq:3}
\end{equation}
since this operator can connect states with different orbital angular momenta,
it can directly induce \(P\)-\(F\) mixing in charmonium.

Relativistic effects in the GI model are incorporated through a smearing
procedure and momentum-dependent correction factors. The smearing function is
introduced as
\begin{equation}
\rho_{ij}(\mathbf r-\mathbf r')
=
\frac{\sigma_{ij}^3}{\pi^{3/2}}
\exp\left[-\sigma_{ij}^2(\mathbf r-\mathbf r')^2\right] .
\end{equation}
With this function, the confining potential
\begin{equation}
S(r)=br+c
\end{equation}
and the one-gluon-exchange potential
\begin{equation}
G(r)=-\frac{4\alpha_s(r)}{3r}
\end{equation}
are smeared according to
\begin{equation}
\widetilde f(r)
=
\int d^3\mathbf r'\,
\rho_{ij}(\mathbf r-\mathbf r') f(r') ,
\qquad
f(r)=G(r),\,S(r).
\end{equation}

Furthermore, the smeared potential is modified by momentum-dependent
relativistic factors. For a general interaction term \(\widetilde V_i(r)\), one
makes the replacement
\begin{equation}
\widetilde V_i(r)
\rightarrow
\left(
\frac{m_i m_j}{E_i E_j}
\right)^{1/2+\epsilon_i}
\widetilde V_i(r)
\left(
\frac{m_i m_j}{E_i E_j}
\right)^{1/2+\epsilon_i},
\end{equation}
where
\begin{equation}
E_i=\left(\mathbf p^2+m_i^2\right)^{1/2},
\qquad
E_j=\left(\mathbf p^2+m_j^2\right)^{1/2},
\end{equation}
the parameter \(\epsilon_i\) depends on the type of interaction considered.
Further details of the GI model can be found in Ref.~\cite{PhysRevD.32.189}.

The spatial wave functions of the mesons, required as input for subsequent decay calculations, are obtained by solving the potential model. We represent the numerical wave function in momentum space as an expansion
\begin{equation}
\Psi_{nLM_{L}}(\mathbf{P})=\sum_{n_{1}}^{n_{\text{max}}}C_{n}R_{nL}(\mathbf{P})Y_{LM_{L}}(\Omega_{\mathbf{P}}), 
\label{eq:4}
\end{equation}
where $R_{nL}(\mathbf{P})$ is the radial wave function and $n_{\text{max}}$ is the size of the basis. We take $n_{\text{max}}=20$, which gives an accurate description of the numerically obtained wave functions. This representation provides the necessary framework for computing OZI‑allowed two‑body strong decay widths.

The parameters of the GI model are fitted to the measured charmonium spectrum. In Ref.~\cite{Duan:2020tsx} a satisfactory description of the low‑lying charmonium masses was achieved, and we therefore adopt the same parameter set here. The values are listed in Table~\ref{tab:gi_model}; for further details on the fitting procedure we refer to Ref.~\cite{Duan:2020tsx}.

\begin{table}[htbp]
    \centering
    \renewcommand{\arraystretch}{1.4}
    
    \caption{The parameters of the GI model.}
    
    \begin{tabular*}{\linewidth}{@{\extracolsep{\fill}}cccc}
        \toprule[1.00pt]
        \toprule[1.00pt]
        Parameters & Values & Parameters & Values \\
        \hline
        $m_q$ & 0.220 GeV & $b$ & 0.175 GeV$^2$ \\
        $m_c$ & 1.628 GeV & $c$ & $-0.245$ GeV \\
        $m_s$ & 0.419 GeV & $\epsilon_{\text{cont}}$ & $-0.103$ \\
        $s$   & 0.821 GeV & $\epsilon_{\text{so(v)}}$ & $-0.279$ \\
        $\sigma_0$ & 2.33 GeV & $\epsilon_{\text{so(s)}}$ & $-0.3$ \\
        $\alpha_s$ & 0.6 & $\epsilon_{\text{tens}}$ & $-0.114$ \\
        $\Lambda$ & 0.2 GeV & & \\
        \bottomrule[1.00pt]
        \bottomrule[1.00pt]
    \end{tabular*}
    \label{tab:gi_model}
\end{table}

Using the parameters in Table~\ref{tab:gi_model}, the GI model yields the bare masses of the charmonia of interest: $\chi_{c2}(2P)$ at 3975.0~MeV and $\chi_{c2}(1F)$ at 4079.0~MeV. The resulting $\chi_{c2}(2P)$ mass is about 50~MeV higher than the mass of the experimentally observed $\chi_{c2}(3930)$. This discrepancy is resolved after incorporating coupled-channel effects and $2P$-$1F$ mixing, which shift the bare mass downward and bring it into agreement with the experimental value.

Similar to the $S$-$D$ mixing scheme, the following relation holds:
\begin{equation}
\begin{pmatrix}
\left|\chi^{\prime}_{c2}\right\rangle \\
\left|\chi^{\prime\prime}_{c2}\right\rangle
\end{pmatrix}
=
\begin{pmatrix}
\cos\theta & -\sin\theta \\
\sin\theta & \cos\theta
\end{pmatrix}
\begin{pmatrix}
|\chi_{c2}(2P)\rangle \\
|\chi_{c2}(1F)\rangle
\end{pmatrix},
\label{eq:5}
\end{equation}
where $\left|\chi^{\prime}_{c2}\right\rangle$ and $\left|\chi^{\prime\prime}_{c2}\right\rangle$ denote the lower- and higher-mass $2P$–$1F$ mixed states, respectively. $\theta$ is the corresponding mixing angle. The lower state $\left|\chi^{\prime}_{c2}\right\rangle$ corresponds to the $\chi_{c2}(3930)$, while the higher state $\left|\chi^{\prime\prime}_{c2}\right\rangle$ can be identified as a charmonium state yet to be discovered.

Starting from the GI model, the tensor term can contribute to the mixing of $2P$- and $1F$-wave charmonium states. In the following, we examine the contribution of the tensor term Eq. (\ref{eq:3}) to the mixing of $2P$- and $1F$-wave charmonium states by solving the equation:
\begin{equation}
\scalebox{0.95}{$\displaystyle
\begin{pmatrix}
m_{P}^{0} & \langle \chi_{c2}(1F)|H_{\mathrm{T}}|\chi_{c2}(2P)\rangle \\
\langle \chi_{c2}(2P)|H_{\mathrm{T}}|\chi_{c2}(1F)\rangle & m_{F}^{0}
\end{pmatrix}
\begin{pmatrix}
C_{P} \\
C_{F}
\end{pmatrix}
= M
\begin{pmatrix}
C_{P} \\
C_{F}
\end{pmatrix}.
$}
\label{eq:6}
\end{equation}

The diagonal terms, $m_{P}^{0}$ and $m_{F}^{0}$, represent the bare masses of $\chi_{c2}(2P)$ and $\chi_{c2}(1F)$, respectively. According to our calculations, the $2P$-$1F$ mixing angle induced by the tensor term is only $0.3^\circ$, which has negligible impact on both the masses and decay properties of $\chi_{c2}(2P)$ and $\chi_{c2}(1F)$ and consistent with Refs.~\cite{Kalashnikova:2005ui,Lu:2016mbb}. This is why we need to find an alternative source for the $2P-1F$ mixing scheme, which will be the main focus of the following section.

Among the newly observed hadronic states, a universal phenomenon known as the "low mass puzzle" has been identified in the X(3872)~\cite{Belle:2003nnu}, $D_{s0}$(2317)~\cite{BaBar:2003oey}, $D_{s1}$(2460)~\cite{CLEO:2003ggt}, and $\Lambda_{c}$(2940)~\cite{BaBar:2006itc}. Specifically, the measured masses of these states are consistently lower than the masses predicted by quenched models when these states are treated as conventional hadrons (i.e., $q\bar{q}$ mesons or $qqq$ baryons). To address this low mass puzzle, a direct approach has been to consider the possibility of exotic hadronic configurations for these states. This approach has opened a new window into the study of hadron spectroscopy over the past few decades~\cite{Chen:2016qju,Guo:2017jvc,Liu:2019zoy,Chen:2022asf,Liu:2024uxn}.

In fact, the emergence of this low mass puzzle has highlighted the limitations of the quenched model, which was once the cornerstone of hadron spectroscopy. The influence of this type of model has persisted to the present day, but the observed discrepancies have prompted a reevaluation of its applicability.

When unquenched effects are taken into account, the low mass puzzle observed in the $X(3872)$, $D_{s0}$(2317), $D_{s1}$(2460), and $\Lambda_{c}$(2940) can be significantly alleviated~\cite{Li:2009zu,vanBeveren:2003kd,vanBeveren:2003jv,Kalashnikova:2005ui,Luo:2019qkm,Man:2024mvl}. This demonstrates that unquenched effects cannot be ignored, and in fact, should be emphasized in modern hadron spectroscopy. It is for this reason that the current era of hadron spectroscopy is increasingly defined by the unquenched model, marking a significant shift in our understanding of hadronic states.

In the study of hadron spectroscopy, there are different approaches to account for the unquenched effect. As demonstrated in previous studies~\cite{Wang:2019mhs,Wang:2018rjg}, one method involves introducing a screening potential within the framework of the potential model. Although this is a phenomenological approach, it effectively captures the realistic unquenched effect, as evidenced by recent applications in the study of higher-lying charmonium states~\cite{Wang:2018rjg,Wang:2020prx}. Another approach is to perform a comprehensive coupled-channel analysis, which explicitly incorporates the coupling between the bare state and its allowed hadronic channels~\cite{Lu:2016mbb,Fu:2018yxq}. The latter method is employed in the present work. It is worth noting that the equivalence of these two treatments has been approximately established in Refs.~\cite{Li:2009ad,Duan:2021alw}.

To account for the coupled-channel effects and potential state mixing, the total Hamiltonian of the system is constructed as the sum of a bare Hamiltonian $H_0$, a continuum Hamiltonian $H_{BC}$, and an interaction term $H_I$:
\begin{equation}
    H = H_0 + H_{BC} + H_I.
\label{eq:8}
\end{equation}
The bare Hamiltonian $H_0$ describes the discrete quarkonium states (the bare core) with masses $m_{P}^{0}$ and $m_{F}^{0}$, corresponding to the $|\chi_{c2}(2P)\rangle$ and $|\chi_{c2}(1F)\rangle$ components, respectively. Additionally, $H_0$ includes a tensor interaction part, $H_T$, which is responsible for the direct mixing between the $P$-wave and $F$-wave components in the absence of continuum coupling. The continuum Hamiltonian $H_{BC}$ governs the dynamics of the meson pairs $|BC; p\rangle$ with total energy $E_{BC}(p)$. Note that we neglect the direct interactions between the mesons in the continuum, treating them as free particles. The interaction Hamiltonian $H_I$ represents the coupling between the discrete bare states and the continuum channels, facilitating the transitions $|\chi_{c2}(2P/1F)\rangle \leftrightarrow |BC\rangle$. To describe the coupling between the bare state and the hadronic channel composed of $ B $ and $ C $, we express the matrix element as $ \langle \chi_{c2}(2P/1F) | H_I |BC, p  \rangle = \mathcal{M}_{SL}(A \to B + C) $. To calculate the amplitude $ \mathcal{M}_{SL}(A \to B + C) $, we employ the quark pair creation (QPC) model~\cite{PhysRevD.8.2223,PhysRevD.9.1415}. The corresponding Hamiltonian is given by:
\begin{align}
H_{I} &= -3\gamma\sum_{m}\langle 1m1-m\mid 00\rangle\int d\mathbf{p}_{3}d\mathbf{p}_{4}\delta^{3}(\mathbf{p}_{3}+\mathbf{p}_{4}) \nonumber \\
&\quad \times\mathcal{Y}_{1}^{m}\left(\frac{\mathbf{p}_{3}-\mathbf{p}_{4}}{2}\right)\chi_{1-m}^{34}\varphi_{0}^{34}\omega_{0}^{34}b_{3}^{\dagger}(\mathbf{p}_{3})d_{4}^{\dagger}(\mathbf{p}_{4}),
\label{eq:9}
\end{align}
where the dimensionless parameter $ \gamma $ represents the strength of the $ q\bar{q} $ pair creation from the vacuum.
If a strange quark pair is created, the effective strength is adjusted to $ \gamma_{s}=\frac{m_{n}}{m_{s}}\gamma $, where $ m_{n}(n=u,d) $ and $ m_{s} $ denote the constituent masses of the nonstrange and strange quarks, respectively. The momenta of the created quark $ q $ and antiquark $ \bar{q} $ are denoted by $ \mathbf{p}_{3} $ and $ \mathbf{p}_{4} $, respectively. The solid harmonic function is given by $ \mathcal{Y}_{1}^{m}(\mathbf{p}) = |\mathbf{p}|Y_{1}^{m}(\theta,\phi) $, and $ \chi_{1-m}^{34} $, $ \varphi_{0}^{34} $, and $ \omega_{0}^{34} $ represent the spin, flavor, and color wave functions of the created $ q\bar{q} $ pair, respectively.

The physical state $|A\rangle$ is then an eigenstate of the total Hamiltonian with eigenvalue $M$, satisfying the Schrödinger equation $H|A\rangle = M|A\rangle$. Using the relations above,  we can obtain the reduced matrix:
\begin{equation}
\scalebox{0.83}{$\displaystyle
\begin{aligned}
\begin{pmatrix}
m_{P}^{0} + \Delta_{P}(M) & \langle \chi_{c2}(2P) \mid H_{\mathrm{T}} \mid \chi_{c2}(1F) \rangle + \Delta_{\mathrm{MIX}}(M) \\
\langle \chi_{c2}(1F) \mid H_{\mathrm{T}} \mid \chi_{c2}(2P) \rangle + \Delta_{\mathrm{MIX}}(M) & m_{F}^{0} + \Delta_{F}(M)
\end{pmatrix}
\end{aligned},
\label{eq:10}
$}
\end{equation}
where $M$ is obtained by solving the eigenvalue equation of the reduced matrix. The masses after mixing and the mixing angle can be obtained by solving for the eigenvalues and eigenstates of the reduced matrix.

In principle, all possible open-charm meson loops should be included in the self-energy function. However, a challenge arises in calculating the infinite number of hadron loops in $\Delta_{P(F)}(M)$ and $\Delta_{\mathrm{MIX}}(M)$. This issue was addressed in Ref.~\cite{Pennington:2007xr}, where the authors proposed the once-subtracted dispersion relation, which effectively limits the number of loops and resolves this problem. 
Here, the same set of hadronic loops is included for the two mixed charmonium
states when evaluating their mass shifts. Specifically, we retain the allowed
open-charm channels whose thresholds lie below the bare masses of
\(\chi_{c2}(2P)\) and \(\chi_{c2}(1F)\).

We now employ the once-subtracted method~\cite{Pennington:2007xr} to write $\Delta_{P(F)}(M)$ and $\Delta_{\mathrm{MIX}}(M)$ as:
\begin{equation}
\label{eq:11}
\begin{aligned}[b]
\begin{split}
\Delta_{P(F)}(M) = \operatorname{Re} \sum_{BC} \int_{0}^{\infty} \frac{\left(M_{J/\psi} - M\right) \left| \mathcal{M}_{P(F)}(p) \right|^{2} }
{\left(M - E_{BC}\right)\left(M_{J/\psi} - E_{BC}\right)}p^{2} \, dp,
\end{split}
\end{aligned}
\end{equation}
and
\begin{equation}
\begin{aligned}[b]
\begin{split}
\Delta_{\mathrm{MIX}}(M)= \operatorname{Re} \sum_{BC} \int_{0}^{\infty} \frac{\left(M_{J/\psi} - M\right) \mathcal{M}_{P}(p) \, \mathcal{M}_{F}^{*}(p)}{\left(M - E_{BC}\right) \left(M_{J/\psi} - E_{BC}\right)} p^{2} \, dp, 
\end{split}
\end{aligned}
\label{eq:12}
\end{equation}
respectively. Here, $\mathcal{M}_{P(F)}(p)$ represents the matrix element for the transition of a $P$-wave or $F$-wave charmonium state to an intermediate mesonic state via the interaction Hamiltonian $\langle \chi_{c2}(2P/1F) | H_I |BC, p  \rangle$. $M_{J/\psi}$ denotes the subtraction point, which we take as the mass of the $J/\psi$ meson in the charmonium system. 

In practical 
unquenched calculations, the collective contributions from the infinite tower of 
higher excited hadron loops can be non-negligible in a conventional un-subtracted 
framework~\cite{Lu:2017hma}. However, within the once-subtracted dispersion relation employed in 
Eq.~(\ref{eq:11}) and Eq.~(\ref{eq:12}), such effects are systematically controlled and compensated for. For deeply 
closed channels whose thresholds $E_{BC}$ are far above the charmonium mass $M$ of 
interest ($E_{BC} \gg M$), their self-energies vary extremely slowly and behave as 
constants. Through the subtraction at the reference point $M_{J/\psi}$, these constant 
real-part contributions naturally cancel out, i.e., $\Pi_n(M) - \Pi_n(M_{J/\psi}) \approx 0$. 
Physically, these subtracted constant  do not vanish but are implicitly 
absorbed into the parameters and bare masses of the quenched potential model, which were originally determined by fitting the physical 
spectrum that already embodies full vacuum-polarization effects. Furthermore, the 
once-subtracted framework introduces an additional energy factor in the denominator 
of Eq.~(\ref{eq:11}) and Eq.~(\ref{eq:12}). This factor ensures that the remaining dynamical contributions from 
higher excited channels are suppressed by a factor of $\sim 1/E_{BC}^2$ at high energies. 
Consequently, the mass shifts are predominantly governed by the low-lying channels 
that are fully open or near the threshold. Studies using this method have successfully reproduced reasonable meson masses, as demonstrated in Refs.~\cite{Duan:2020tsx,Duan:2021alw,Ni:2023lvx,Deng:2023mza}.

Using the amplitude $ \mathcal{M}_{SL}(A \to B + C) $, the OZI-allowed two-body strong decay width for the charmonium state can be calculated as:
\begin{equation}
\Gamma_{\text{total}}=\sum_{BC}\frac{2\pi PE_{B}E_{C}}{M}\sum_{JL}|\mathcal{M}_{SL}(A \to B + C)|^{2}, 
\label{eq:13}
\end{equation}
where $ P $ is the momentum of the final-state mesons, and $ M $ is the mass of the physical state under consideration.

\section{Numerical results}
\label{sec:numerical}

Before discussing the $2P$-$1F$ mixing induced by coupled-channel effects, it is essential to quantitatively investigate how these channels affect the bare charmonium states $\chi_{c2}(2P)$ and $\chi_{c2}(1F)$. This analysis clarifies the origin and magnitude of their mass shifts. For $\chi_{c2}(2P)$ and $\chi_{c2}(1F)$, the coupled-channel equation~(\ref{eq:10}) can be written as
\begin{equation}
M-m_{P(F)}^{0}-\Delta_{P(F)}(M)=0,
\label{eq:14}
\end{equation}
where $\Delta_{P(F)}(M)$ denotes the mass shift from open-charm channels.

The calculation of the mass shifts $\Delta_{P(F)}(M)$ relies on the computation of the transition amplitudes $ \mathcal{M}_{P(F)}(p)$ in the QPC model, which in turn depends on the parameter $\gamma$. In our calculations, we adopt $ \gamma = 0.40 $ as the strength parameter for the quark pair creation. 
This value is chosen because it reproduces the decay widths well of $\psi(3770)$ and $\psi(4040)$~\cite{Duan:2020tsx}. If the parameter $\gamma$ deviates too far from $0.40$, the calculated mass and width would become less reliable. To investigate its influence on the final results, we perform calculations with $\gamma = 0.37$, $0.40$, and $0.43$. The central values of our results correspond to $\gamma = 0.40$, while the superscripts and subscripts represent the deviations obtained with $\gamma = 0.37$ and $\gamma = 0.43$, respectively.

Using the parameters described earlier, Table~\ref{tab:combined_mass_shifts} summarizes the resulting masses, mass shifts, and decay widths. 
The coupled-channel dressed masses of $\chi_{c2}(2P)$ and $\chi_{c2}(1F)$ are taken as $3920.9^{+6.7}_{-6.8}$~MeV and $4020.6^{+9.3}_{-10.1}$~MeV, respectively. 
After including coupled-channel effects, $\chi_{c2}(2P)$ acquires a downward mass shift of $-54.1^{+6.7}_{-6.8}$~MeV, dominated by its coupling to the $DD^{*}$ channel in a $D$-wave. The $DD$ and $D_{s}D_{s}$ channels give smaller contributions, also via $D$-wave couplings. Similarly, $\chi_{c2}(1F)$ receives a mass shift of $-58.5^{+9.4}_{-10.1}$~MeV, again driven predominantly by $DD^{*}$ ($D$-wave), with minor additions from $DD$ and $D_{s}D_{s}$ channels.

\begin{table*}[htbp]
\caption{The mass shifts ($\Delta M_i$) and decay widths ($\Gamma_i$) of the $\chi_{c2}(2P)$ and $\chi_{c2}(1F)$ states. The values are given in units of MeV. }
\label{tab:combined_mass_shifts}
\renewcommand{\arraystretch}{1.5}

\begin{tabular*}{\textwidth}{@{\extracolsep{\fill}}lcccccccc}
    \toprule[1.00pt]
    \toprule[1.00pt]
    & \multicolumn{4}{c}{$\chi_{c2}(2P)$} & \multicolumn{4}{c}{$\chi_{c2}(1F)$} \\
    \cmidrule(lr){2-5} \cmidrule(lr){6-9}
    Channel & $\Delta M_i$ & $\Delta M_i / \Delta M_{\text{total}}$ & $\Gamma_i$ & $\Gamma_i / \Gamma_{\text{total}}$ & $\Delta M_i$ & $\Delta M_i / \Delta M_{\text{total}}$ & $\Gamma_i$ & $\Gamma_i / \Gamma_{\text{total}}$ \\
    \hline
    $D\bar{D}$        & $-11.3^{+1.9}_{-2.1}$ & $20.8^{+1.1}_{-1.0}\%$ & $21.8^{+2.3}_{-2.5}$ & $74.0^{+5.7}_{-5.3}\%$ & $-16.6^{+3.3}_{-4.0}$ & $28.5^{+1.5}_{-1.5}\%$ & $68.0^{+8.1}_{-8.1}$  & $58.9^{+2.2}_{-1.9}\%$\\
    $D\bar{D}^*$      & $-40.9^{+4.6}_{-4.5}$ & $75.6^{+1.1}_{-1.1}\%$ & $7.6^{+1.2}_{-1.5}$  & $26.0^{+5.3}_{-5.7}\%$ & $-38.7^{+5.6}_{-5.8}$ & $66.2^{+1.1}_{-1.3}\%$ & $46.4^{+1.1}_{-2.4}$  & $40.2^{+1.7}_{-2.0}\%$\\
    $D_s\bar{D}_s$   & $-1.9^{+0.2}_{-0.3}$  & $3.6^{+0.0}_{-0.1}\%$  & --   & --     & $-3.1^{+0.3}_{-0.4}$  & $5.3^{+0.4}_{-0.3}\%$  & $1.1^{+0.0}_{-0.2}$   & $0.9^{+0.2}_{-0.1}\%$\\ 
    \hline
    Total             & $-54.1^{+6.7}_{-6.8}$ & 100\%  & $29.4^{+0.8}_{-1.2}$ & 100\%  & $-58.5^{+9.4}_{-10.1}$ & 100\%  & $115.6^{+8.9}_{-10.6}$ & 100\% \\
    \hline
    $M_{\text{Th}}/\Gamma_{\text{Th}}$   & \multicolumn{2}{c}{$3920.9^{+6.7}_{-6.8}$} & \multicolumn{2}{c}{$29.4^{+0.8}_{-1.2}$} & \multicolumn{2}{c}{$4020.6^{+9.3}_{-10.1}$} & \multicolumn{2}{c}{$115.6^{+8.9}_{-10.6}$} \\
    $M_{\text{Exp}}/\Gamma_{\text{Exp}}$~\cite{ParticleDataGroup:2024cfk} & \multicolumn{2}{c}{$3922.5^{+1.0}_{-1.0}$} & \multicolumn{2}{c}{$35.2^{+2.2}_{-2.2}$} & \multicolumn{2}{c}{--}     & \multicolumn{2}{c}{--} \\
    \bottomrule[1.00pt]
    \bottomrule[1.00pt]
\end{tabular*}
\end{table*}

By solving Eq.~(\ref{eq:10}) within our framework, we obtain the physical masses of the mixed states $\chi^{\prime}_{c2}$ and $\chi^{\prime\prime}_{c2}$ as $3919.3^{+6.8}_{-6.8}$~MeV and $4030.0^{+6.9}_{-7.6}$~MeV, respectively.
The corresponding eigenvectors are listed in Table~\ref{tab:mixing_angles}, where the components $(C_{P}^{\prime/\prime\prime}, C_{F}^{\prime/\prime\prime})$ denote the mixing coefficients of the $\chi_{c2}(2P)$ and $\chi_{c2}(1F)$ bare bases in each physical state. From these coefficients, the mixing angles are determined to be
\begin{equation}
\theta_{1} =\arctan(-C_{F}^{\prime}/C_{P}^{\prime})  = (7.5^{-0.3}_{+0.2})^\circ \quad \text{for} \quad \chi^{\prime}_{c2},
\label{eq:15}
\end{equation}
and
\begin{equation}
\theta_{2} =\arctan(C_{P}^{\prime\prime}/C_{F}^{\prime\prime}) = (15.4^{-1.9}_{+1.7})^{\circ} \quad \text{for} \quad \chi^{\prime\prime}_{c2},
\label{eq:16}
\end{equation}
within our framework, the mixing is mainly induced by the coupled-channel effects rather than the tensor force in the potential model, which is consistent with Refs.~\cite{Man:2025zfu,Man:2025vmm,Lu:2016mbb}.

In the conventional \(S\)-\(D\) mixing analyses, the mixing angle is usually
extracted from the leptonic widths of \(\psi(3686)\) and \(\psi(3770)\)~\cite{Rosner:2001nm,Kuang:1989ub}. Since
the leptonic width is proportional to the squared modulus of the amplitude,
\begin{equation}
\Gamma_{ee}\propto |\mathcal A_{ee}|^2 ,
\end{equation}
the experimental input determines only the magnitude of the amplitude ratio, 
leaving its relative sign undetermined.
As a result, two solutions with opposite signs of the mixing angle can reproduce the same leptonic-width measurements.
Therefore, the opposite-sign angles quoted in the literature are a consequence of the specific extraction procedure employed in Refs.~\cite{Rosner:2001nm,Kuang:1989ub}, rather than a universal requirement of state mixing.

In our coupled-channel framework, the mass shifts and off-diagonal mixing
terms depend explicitly on the physical mass through the self-energy corrections.
Consequently, the effective Hamiltonian appearing in the eigenvalue equation differs for distinct physical solutions, i.e.,
\begin{equation}
H(M_1) \neq H(M_2).
\end{equation}

The problem is thus a nonlinear eigenvalue problem rather than the diagonalization
of a single fixed matrix. As a result, the two physical states are not required to be eigenvectors of the same
Hamiltonian and need not be strictly orthogonal. Therefore, they cannot generally
be described by a single mixing angle. For this reason, we introduce two mixing
angles corresponding to the two physical solutions.The fact that both extracted
angles shown in Eq.~(\ref{eq:15}) and Eq.~(\ref{eq:16}) are positive is a consequence of our phase convention for the eigenvectors listed in Table~\ref{tab:mixing_angles}.

\begin{table}[htbp] 
    \centering
    \caption{The $2P$-$1F$ mixing angles for $\chi^{\prime}_{c2}$ and $\chi^{\prime\prime}_{c2}$ are induced by the coupled-channel effects. The mixing states are expressed as $C_{P}^{\prime/\prime\prime}|\chi_{c2}(2P)\rangle + C_{F}^{\prime/\prime\prime}|\chi_{c2}(1F)\rangle$.} 
    \label{tab:mixing_angles}

    \renewcommand{\arraystretch}{1.5}
    
    \begin{tabular*}{\columnwidth}{@{\extracolsep{\fill}}lcccc}
        \toprule[1.00pt]
        \toprule[1.00pt]
        States & $C_{P}^{\prime/\prime\prime}$ & $C_{F}^{\prime/\prime\prime}$ & Mixing angle & Mass (MeV) \\
        \hline
        $\chi^{\prime}_{c2}$ & $-0.991^{-0.001}_{+0.000}$ & $0.131^{-0.006}_{+0.003}$ & $\theta_1 = (7.5^{-0.3}_{+0.2})^\circ$ & $3919.3^{+6.8}_{-6.8}$ \\
        $\chi^{\prime\prime}_{c2}$ & $0.265^{-0.031}_{+0.030}$ & $0.964^{+0.008}_{-0.008}$ & $\theta_2 = (15.4^{-1.9}_{+1.7})^{\circ}$ & $4030.0^{+6.9}_{-7.6}$ \\
        \bottomrule[1.00pt]
        \bottomrule[1.00pt]
    \end{tabular*}
\end{table}

Beyond the mass spectrum, we have also investigated the OZI-allowed two-body strong decays of the mixed states. The total decay width of $\chi^{\prime}_{c2}$ is predicted to be 22.7 MeV, in good agreement with the measured width of $\chi_{c2}(3930)$. Its dominant decay mode is $D\bar{D}$, with a branching fraction of $62.9\%$. These results show that a consistent coupled-channel description can accommodate a sizable $2P$-$1F$ mixing angle while simultaneously reproducing the mass and width of $\chi_{c2}(3930)$. This supports the interpretation of $\chi_{c2}(3930)$ as the lower mixed state $\chi^{\prime}_{c2}$.

For the higher mixed state $\chi^{\prime\prime}_{c2}$, the total width is predicted to be 113.2 MeV, also dominated by the $D\bar{D}$ channel with a branching fraction of $79.5\%$. Observing this state experimentally will require high-precision measurements and a dedicated analysis of the $D\bar{D}$ invariant-mass spectrum.

\begin{table}[htbp]
\caption{Two-body strong decay widths of $\chi^{\prime}_{c2}$ and $\chi^{\prime\prime}_{c2}$ with mixing angles $7.5^\circ$ and $15.4^\circ$, respectively. The decay widths are in units of MeV.}
    \label{tab:mixing_decay}

\renewcommand{\arraystretch}{1.5}

\begin{tabular*}{\linewidth}{@{\extracolsep{\fill}}lcccc}
    \toprule[1.00pt]
    \toprule[1.00pt]
    & \multicolumn{2}{c}{$\chi^{\prime}_{c2}$} & \multicolumn{2}{c}{$\chi^{\prime\prime}_{c2}$} \\
    & \multicolumn{2}{c}{$\theta_1 = (7.5^{-0.3}_{+0.2})^\circ$} & \multicolumn{2}{c}{$\theta_2 = (15.4^{-1.9}_{+1.7})^{\circ}$} \\
    \hline
    Channels & $\Gamma_i$ & $\Gamma_i / \Gamma_{\text{total}}$ & $\Gamma_i$ & $\Gamma_i / \Gamma_{\text{total}}$ \\
    \hline
    $D\bar{D}$      & $14.3^{+0.8}_{-0.8}$ & $62.9^{-6.0}_{+7.1}$\% & $90.0^{-0.5}_{-0.1}$ & $79.5^{-3.7}_{+3.4}$\% \\
    $D\bar{D}^*$    & $8.4^{+3.1}_{+2.6}$  & $37.1^{+6.0}_{-7.1}$\% & $21.3^{+5.1}_{-4.3}$ & $18.8^{+3.6}_{-3.1}$\% \\
    $D_s\bar{D}_s$  & --   & --     & $1.9^{+0.2}_{-0.3}$  & $1.7^{+0.1}_{-0.3}$\%  \\
    \hline
    Total           & $22.7^{+3.9}_{-3.7}$ & 100\%  & $113.2^{+4.9}_{-4.6}$ & 100\% \\
    \bottomrule[1.00pt]
    \bottomrule[1.00pt]
\end{tabular*}
\end{table}

It is meaningful to calculate the two-photon and two-gluon widths of the mixed states using the wave function at the origin.

The two-photon decay widths of the mixed tensor charmonia are evaluated in the leading-order nonrelativistic quark model~\cite{PhysRevD.32.189,PhysRevD.37.3210,PhysRevD.46.2257,Robinett:1992px}. In this framework, the decay process is factorized into a short-distance $c\bar c\to\gamma\gamma$ amplitude and the spatial wave function of the charmonium state at short distances. Consequently, the leading contribution from a $P$-wave component is proportional to the first derivative of the radial wave function at the origin, $R'_{2P}(0)$, whereas that from an $F$-wave component is proportional to the third derivative, $R'''_{1F}(0)$. For a pure $2\,{}^3P_2$ state, the dominant two-photon helicity amplitude is the $\lambda=2$ component, which gives
\begin{equation}
\Gamma_{\gamma\gamma}\left(2\,{}^3P_2\right)
=
\frac{36}{5}\,
\alpha^2 e_c^4
\frac{|R'_{2P}(0)|^2}{m_c^4}.
\end{equation}
For a pure $1\,{}^3F_2$ state, both $\lambda=2$ and $\lambda=0$ two-photon helicity amplitudes contribute at leading order. Following the standard nonrelativistic helicity-amplitude results for $F$-wave quarkonia, these contributions can be written as
\begin{equation}
\Gamma_{\gamma\gamma}^{\lambda=2}\left(1\,{}^3F_2\right)
=
\frac{125}{6}\,
\alpha^2 e_c^4
\frac{|R'''_{1F}(0)|^2}{m_c^8},
\end{equation}
and
\begin{equation}
\Gamma_{\gamma\gamma}^{\lambda=0}\left(1\,{}^3F_2\right)
=
\frac{49}{5}\,
\alpha^2 e_c^4
\frac{|R'''_{1F}(0)|^2}{m_c^8}.
\end{equation}
Here, $\alpha$ is the fine-structure constant, $e_c=2/3$ is the charm-quark electric charge in units of $|e|$, and $m_c$ is the charm-quark mass used in the potential model.

For the physical mixed states,
\begin{equation}
\begin{pmatrix}
|\chi'_{c2}\rangle \\
|\chi''_{c2}\rangle
\end{pmatrix}
=
\begin{pmatrix}
\cos\theta & -\sin\theta \\
\sin\theta & \cos\theta
\end{pmatrix}
\begin{pmatrix}
|2\,{}^3P_2\rangle \\
|1\,{}^3F_2\rangle
\end{pmatrix},
\end{equation}
the two-photon decay amplitudes are obtained by coherently adding the $P$- and $F$-wave annihilation amplitudes. Therefore, for the lower mixed state one has
\begin{equation}
\Gamma_{\gamma\gamma}^{\lambda=2}(\chi'_{c2})
=
\left|
\cos\theta_{1}
\sqrt{\frac{36}{5}}\frac{R'_{2P}(0)}{m_c^2}
-
\sin\theta_{1}
\sqrt{\frac{125}{6}}\frac{R'''_{1F}(0)}{m_c^4}
\right|^2
\alpha^2 e_c^4 ,
\end{equation}
\begin{equation}
\Gamma_{\gamma\gamma}^{\lambda=0}(\chi'_{c2})
=
\left|
\sin\theta_{1}
\sqrt{\frac{49}{5}}\frac{R'''_{1F}(0)}{m_c^4}
\right|^2
\alpha^2 e_c^4 ,
\end{equation}
and
\begin{equation}
\Gamma_{\gamma\gamma}(\chi'_{c2})
=
\Gamma_{\gamma\gamma}^{\lambda=2}(\chi'_{c2})
+
\Gamma_{\gamma\gamma}^{\lambda=0}(\chi'_{c2}) .
\end{equation}
Similarly, for the higher mixed state,
\begin{equation}
\begin{split}
\Gamma_{\gamma\gamma}(\chi''_{c2})
&=\left|\sin\theta_{2}\sqrt{\frac{36}{5}}\frac{R_{2P}^{\prime}(0)}{m_{c}^{2}}+\cos\theta_{2}\sqrt{\frac{125}{6}}\frac{R_{1F}^{\prime\prime\prime}(0)}{m_{c}^{4}}\right|^{2}\alpha^{2}e_{c}^{4}\\&\quad+\left|\cos\theta_{2}\sqrt{\frac{49}{5}}\frac{R_{1F}^{\prime\prime\prime}(0)}{m_{c}^{4}}\right|^{2}\alpha^{2}e_{c}^{4},
\end{split}
\end{equation}
the relative signs in these expressions follow from the mixing convention adopted above.

According to Ref.~\cite{Robinett:1992px}, in the nonrelativistic formalism the annihilation amplitude is expressed
 as a short-distance annihilation Dirac operator and a long-distance bound state
wave function. The conversion from the two-photon width to the two-gluon
width therefore corresponds to replacing the short-distance process
$c\bar c\to \gamma\gamma$ by $c\bar c\to gg$, while keeping the same
wave-function. For a color-singlet $c\bar c$ state, this replacement
introduces the corresponding coupling constant and color factor, leading to
\begin{equation}
\Gamma_{gg}
=
\frac{2\alpha_s^2}{9\alpha^2 e_c^4}
\,
\Gamma_{\gamma\gamma}.
\end{equation}
in the present work, we take $\alpha_s=0.332$ for the charm quark and use
this leading-order relation to estimate the two-gluon widths of the mixed
tensor charmonia.

In this work, we employ the GI model to calculate the wave functions, where $R_{2P}^{\prime}(0) = 0.12\ \text{GeV}^{5/2}$ is the first derivative of the $P$-wave radial wave function at the origin, and $R_{1F}^{\prime\prime\prime}(0) = 0.06\ \text{GeV}^{9/2}$ is the corresponding third derivative for the $F$-wave.

With the mixing angle $\theta_1 = (7.5^{-0.3}_{+0.2})^\circ$, we obtain
\[
\Gamma_{\gamma\gamma}(\chi^{\prime}_{c2}) = 0.145^{+0.000}_{-0.001}\ \text{keV}, \qquad
\Gamma_{gg}(\chi^{\prime}_{c2}) = 0.337^{+0.001}_{-0.002}\ \text{MeV}.
\]
For $\theta_2 = (15.4^{-1.9}_{+1.7})^{\circ}$ the corresponding widths are
\[
\Gamma_{\gamma\gamma}(\chi^{\prime\prime}_{c2}) = 0.055^{-0.005}_{+0.004}\ \text{keV}, \qquad
\Gamma_{gg}(\chi^{\prime\prime}_{c2}) = 0.127^{-0.009}_{-0.011}\ \text{MeV}.
\]

The suppression of decay widths for $\chi^{\prime\prime}_{c2}$ arises from its larger $F$-wave component. Since the wave function at the origin (or its higher-order derivative) for an $F$-wave ($L=3$) state is much smaller than that of a $P$-wave state, the resulting transition strengths are significantly reduced. This is reflected in our results, where $\Gamma_{\gamma\gamma}$ decreases from $0.14$ keV for  $\chi^{\prime}_{c2}$ to $0.05$ keV for  $\chi^{\prime\prime}_{c2}$, and $\Gamma_{gg}$ drops from $0.33$ MeV to $0.13$ MeV.

\section{Production of $2P$-$1F$ mixing states via $\gamma\gamma$ fusion}\label{sec4}

In the absence of the $2P$-$1F$ mixing mechanism, the production of the $\chi_{c2}(2P)$ or $\chi_{c2}(1F)$ state via $e^+e^-$ annihilation has been studied in Refs.~\cite{Qian:2023taw,Gao:2024qth,Liu:2024tgq}. Based on this work, we argue that a further analysis of the production properties of our calculated mixed states is warranted. Therefore, in this section, we analyze the productions of $2P$-$1F$ mixed states via $\gamma\gamma$ fusion.


Our study first focuses on the $J/ \psi \omega$ final states, and the total cross section is calculated using Eq.~(\ref{eq:21})
\begin{equation}
\begin{aligned}
\mathcal{\sigma} &\left[ \,
\scalebox{0.45}{
\begin{tikzpicture}[baseline=-0.5ex]

    \draw[thick, decorate, decoration={snake, amplitude=2pt, segment length=7pt, post length=1pt}] 
        (-1.8, 1.2) -- (0,0);
    \draw[thick, decorate, decoration={snake, amplitude=2pt, segment length=7pt, post length=1pt}] 
        (-1.8, -1.2) -- (0,0);
    
    \node at (-2.1, 1.2) {\large $\gamma$};
    \node at (-2.1, -1.2) {\large $\gamma$};

    \draw[line width=1.2pt] (2.5, 0) -- (4.2, 1.2) node[right, xshift=2pt] {\large $J/\psi$};
    \draw[line width=1.2pt] (2.5, 0) -- (4.2, -1.2) node[right, xshift=2pt] {\large $\omega$};

    \draw[thick, double, double distance=1.5pt] (0,0) -- (2.5,0);
    
    \node[above] at (1.25, 0.1) {\large $\chi^{\prime(\prime\prime)}_{c2}$};

    \fill[black] (0,0) circle (0.1);
    \node[fill=gray, minimum size=10pt, inner sep=0pt] at (2.5,0) {};

\end{tikzpicture}
}
\, \right] = \frac{5}{4} \frac{4\pi}{k^2} \frac{s \Gamma_{\mathrm{\gamma \gamma}}(s) \Gamma_{\mathrm{\textit{J}/\psi \omega}}(s)}{(s - M_{\chi^{\prime(\prime\prime)}_{c2}}^2)^2 + s \Gamma_{\mathrm{tot}}^2(s)},
\end{aligned}
\label{eq:21}
\end{equation}
where $s$ is the c.m. energy and the factor of $\frac{5}{4}$ arises from $\frac{2J+1}{(2S_1+1)(2S_2+1)}$, $J$ is the spin of the resonance, and the number of polarization states of the two incident particles are $2S_1+1$ and $2S_2+1$. The c.m. momentum in the initial state is $k$, $M_{\chi^{\prime(\prime\prime)}_{c2}}$ is the mass of the resonance, and $\Gamma_{\mathrm{tot}}$ is the total width of the resonance. $\Gamma_{\mathrm{\gamma \gamma}}$ is the two-photon width and $\Gamma_{\mathrm{\textit{J}/\psi \omega}}$ is the partial width of final states. Under the narrow-width approximation, $s$ can be approximated by the mass of the resonance. Within this cross-section expression, only $\Gamma_{\mathrm{\textit{J}/\psi \omega}}$ is an unknown quantity that requires a dedicated calculation.

For high-lying excited states, coupled-channel effect is significant, so we calculate $\Gamma_{\mathrm{\textit{J}/\psi \omega}}$ using the hadronic loop mechanism~\cite{Bai:2026atm}. In the framework of the hadronic loop mechanism, the initial charmonium transitions into final states via a $D^{(*)}\bar{D}^{(*)}$ intermediate pair and the exchange of a $D^{(*)}$ meson. The decay amplitude is expressed as:
\begin{equation}
\mathcal{M} = \int \frac{d^4 q}{(2\pi)^4} \frac{\mathcal{V}_1 \mathcal{V}_2 \mathcal{V}_3}{\mathcal{G}_1 \mathcal{G}_2 \mathcal{G}_3} \mathcal{F}^2(q^2),
\label{eq:22}
\end{equation}
where $\mathcal{V}_i (i=1,2,3)$ and $\mathcal{G}_i (i=1,2,3)$ denote the interaction vertices and propagators, respectively. 

To account for the off-shell effects of the exchanged meson and regularize the loop integral, a dipole form factor is introduced $
\mathcal{F}(q^2) = \left( \frac{m_E^2 - \Lambda^2}{q^2 - \Lambda^2} \right)^2$,
where $m_E$ and $q$ are the mass and four-momentum of the exchanged charmed meson, respectively. The cut-off parameter is defined as $\Lambda = m_E + \alpha \Lambda_{QCD}$, with $\Lambda_{QCD} = 220$ MeV and $\alpha$ being a dimensionless free parameter of order 1~\cite{Cheng:2004ru}.

In evaluating the coupling vertices, we employ the effective Lagrangian approach, 
which has yielded satisfactory results in numerous previous works~\cite{Gao:2026hjv,Wang:2024xvq,Wang:2024whi}. 
The specific Lagrangians used in our calculation are adopted from Refs.~\cite{Gao:2024qth,Chen:2013yxa}:
\begin{equation}
\begin{aligned}
\mathcal{L}_{\chi^{\prime(\prime\prime)}_{c2} D^{(*)} D^{(*)}} = & \;i g_{\chi^{\prime(\prime\prime)}_{c2} DD} (\chi^{\prime(\prime\prime)}_{c2})^{\mu\nu} \partial_\mu D \partial_\nu D^\dagger\\
& + g_{\chi^{\prime(\prime\prime)}_{c2} DD^*} \varepsilon_{\mu\nu\alpha\beta} \partial^\alpha (\chi^{\prime(\prime\prime)}_{c2})^{\mu\rho} \Bigl( \partial_\rho D^{*\nu} \partial^\beta D^\dagger \\
&- \partial^\beta D \partial_\rho D^{*\nu\dagger} \Bigl) + i g_{\chi^{\prime(\prime\prime)}_{c2} D^* D^*} (\chi^{\prime(\prime\prime)}_{c2})^{\mu\nu} D_\mu^* D_\nu^{*\dagger}  ,
\end{aligned}
\label{eq:24}
\end{equation}

\begin{equation}
\begin{aligned}
\mathcal{L}_{J/\psi D^{(*)}D^{(*)}} = & \; i g_{J/\psi DD} \psi_\mu \left( \partial^\mu D D^\dagger - D \partial^\mu D^\dagger \right) \\
&- g_{J/\psi D^* D} \varepsilon^{\mu\nu\alpha\beta} \partial_\mu \psi_\nu \left( \partial_\alpha D^*_\beta D^\dagger + D \partial_\alpha D^{*\dagger}_\beta \right) \\
& - i g_{J/\psi D^* D^*} \Big\{ \psi^\mu \left( \partial_\mu D^{*\nu} D^{*\dagger}_\nu - D^{*\nu} \partial_\mu D^{*\dagger}_\nu \right)\\ 
&+ \left( \partial_\mu \psi_\nu D^{*\nu} - \psi_\nu \partial_\mu D^{*\nu} \right) D^{*\mu\dagger} \\
& + D^{*\mu} \left( \psi^\nu \partial_\mu D^{*\dagger}_\nu - \partial_\mu \psi_\nu D^{*\nu\dagger} \right) \Big\},
\end{aligned}
\label{eq:25}
\end{equation}

\begin{equation}
\begin{aligned}
\mathcal{L}_{\mathcal{D}^{(*)}\mathcal{D}^{(*)}\mathbb{V}} = & -i g_{\mathcal{DD}\mathbb{V}} \mathcal{D}_i^\dagger \overleftrightarrow{\partial}_\mu \mathcal{D}^j (\mathbb{V}^\mu)^i_j \\
&- 2 f_{\mathcal{D}^* \mathcal{D} \mathbb{V}} \varepsilon_{\mu\nu\alpha\beta} (\partial^\mu \mathbb{V}^\nu)^i_j \left( \mathcal{D}_i^\dagger \overleftrightarrow{\partial}^\alpha \mathcal{D}^{*\beta j}- \mathcal{D}_i^{*\beta\dagger} \overleftrightarrow{\partial}^\alpha \mathcal{D}^j \right) \\
& + i g_{\mathcal{D}^*\mathcal{D}^*\mathbb{V}} \mathcal{D}_i^{*\nu\dagger} \overleftrightarrow{\partial}_\mu \mathcal{D}^{*j}_{\nu} (\mathbb{V}^\mu)^i_j \\
&+4 i f_{\mathcal{D}^*\mathcal{D}^*\mathbb{V}} \mathcal{D}_{i\mu}^{*\dagger} (\partial^\mu \mathbb{V}^\nu - \partial^\nu \mathbb{V}^\mu)^i_j \mathcal{D}_{\nu}^{*j},
\end{aligned}
\label{eq:26}
\end{equation}
where $\mathcal{D} = (D^0, D^+, D_s^+)$, $(\mathcal{D}^\dagger)^T = (\bar{D}^0, D^-, D_s^-)$, and $\overleftrightarrow{\partial} = \overrightarrow{\partial} - \overleftarrow{\partial}$. The light vector nonet meson can form the following $3 \times 3$ matrix $\mathbb{V}$:

\begin{equation}
\scalebox{1.2}{$\displaystyle
\mathbb{V} = 
\begin{pmatrix}
\frac{\rho^0}{\sqrt{2}} + \frac{\omega}{\sqrt{2}} & \rho^+ & K^{*+} \\
\rho^- & \frac{-\rho^0}{\sqrt{2}} + \frac{\omega}{\sqrt{2}} & K^{*0} \\
K^{*-} & \bar{K}^{*0} & \phi
\end{pmatrix}.
$}
\label{eq:27}
\end{equation} 

\begin{table}[htbp]
\centering

\caption{The values of the coupling constants for the charmed mesons coupling to the low-mass ($\chi^{\prime}_{c2}$) and high-mass ($\chi^{\prime\prime}_{c2}$) mixed states.}
\label{tab:coupling_constants_updated}

\renewcommand{\arraystretch}{1.6} 
\begin{tabular*}{\linewidth}{@{\extracolsep{\fill}}lcc}
\toprule[1.00pt]
\toprule[1.00pt]
Coupling Constants & $\chi^{\prime}_{c2}$ & $\chi^{\prime\prime}_{c2}$ \\
\hline

$g_{\chi^{\prime(\prime\prime)}_{c2} DD}$ [$\text{GeV}^{-1}$]      & 17.35 & 24.05 \\
$g_{\chi^{\prime(\prime\prime)}_{c2} DD^*}$ [$\text{GeV}^{-2}$]    & 12.89 & 3.74  \\
\bottomrule[1.00pt]
\bottomrule[1.00pt]
\end{tabular*}
\end{table}

\begin{figure*}[htbp]
	\centering
	\begin{tabular}{c}
		\includegraphics[width=0.98\textwidth]{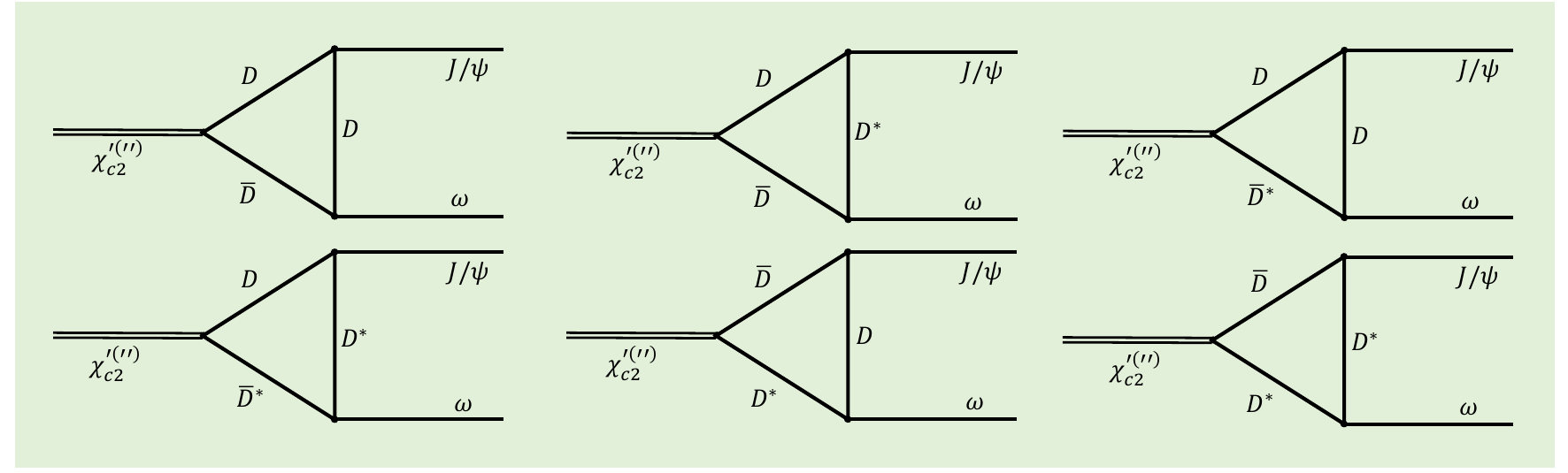}
	\end{tabular}
	\caption{
   The schematic diagrams involving hadronic loops used in the calculation.}
	\label{Fig1}
\end{figure*}

It is important to note that Eq.~(\ref{eq:24}) serves as the Lagrangian for both the low-mass mixed state $\chi^{\prime}_{c2}$ and the high-mass mixed state $\chi^{\prime\prime}_{c2}$, differing only in their coupling constants. It is written in the form of the bare-state Lagrangian for $\chi_{c2}(2P)$. This choice is motivated by the fact that both mixed states share the same quantum numbers, $2^{++}$. Within the framework of effective field theory, we retain the leading-order contributions, while absorbing the effects of higher-order contributions into the corresponding coupling constants. The corresponding coupling constants are listed in Table~\ref{tab:coupling_constants_updated}.
For the charmed mesons coupling to vector meson, we have $g_{\mathcal{DDV}} = g_{\mathcal{D}^*\mathcal{D}^*\mathcal{V}} = \beta g_V / \sqrt{2}$, $f_{\mathcal{D}^*\mathcal{D}\mathcal{V}} = f_{\mathcal{D}^*\mathcal{D}^*\mathcal{V}}/m_{D^*} = \lambda g_V / \sqrt{2}$ with $\beta = 0.9$, $\lambda = 0.56$ GeV$^{-1}$, $g_V = m_\rho / f_\pi$, and $f_\pi = 132$ MeV. In the calculation, the mass and width of the initial state are obtained from our previous calculation results. From the effective Lagrangians, we obtain the Feynman rules for the vertices, from which the amplitudes can be written down. The Feynman diagrams employed in our calculations are shown in Fig.~\ref{Fig1}.

All partial amplitudes and the total amplitude are listed in Appendix~\ref{Amplitudes}.
Using the amplitude we have obtained, the $\chi_{c2}^{\prime(\prime\prime)}\to J/\psi\omega$ decay width can be calculated directly:
\begin{equation}
\Gamma = \frac{1}{2J + 1} \frac{|\vec{p}|}{8\pi m_{\chi^{\prime(\prime\prime)}_{c2}}^2} |\mathcal{M}|^2,
\label{eq:35}
\end{equation}
where $\vec{p}$ represents the center-of-mass momentum of the final state, and the factor of $1/(2J + 1)$ accounts for the averaging over the polarization of the initial state.

\begin{figure*}[htbp]
	\centering
	\begin{tabular}{c}
		\includegraphics[width=0.98\textwidth]{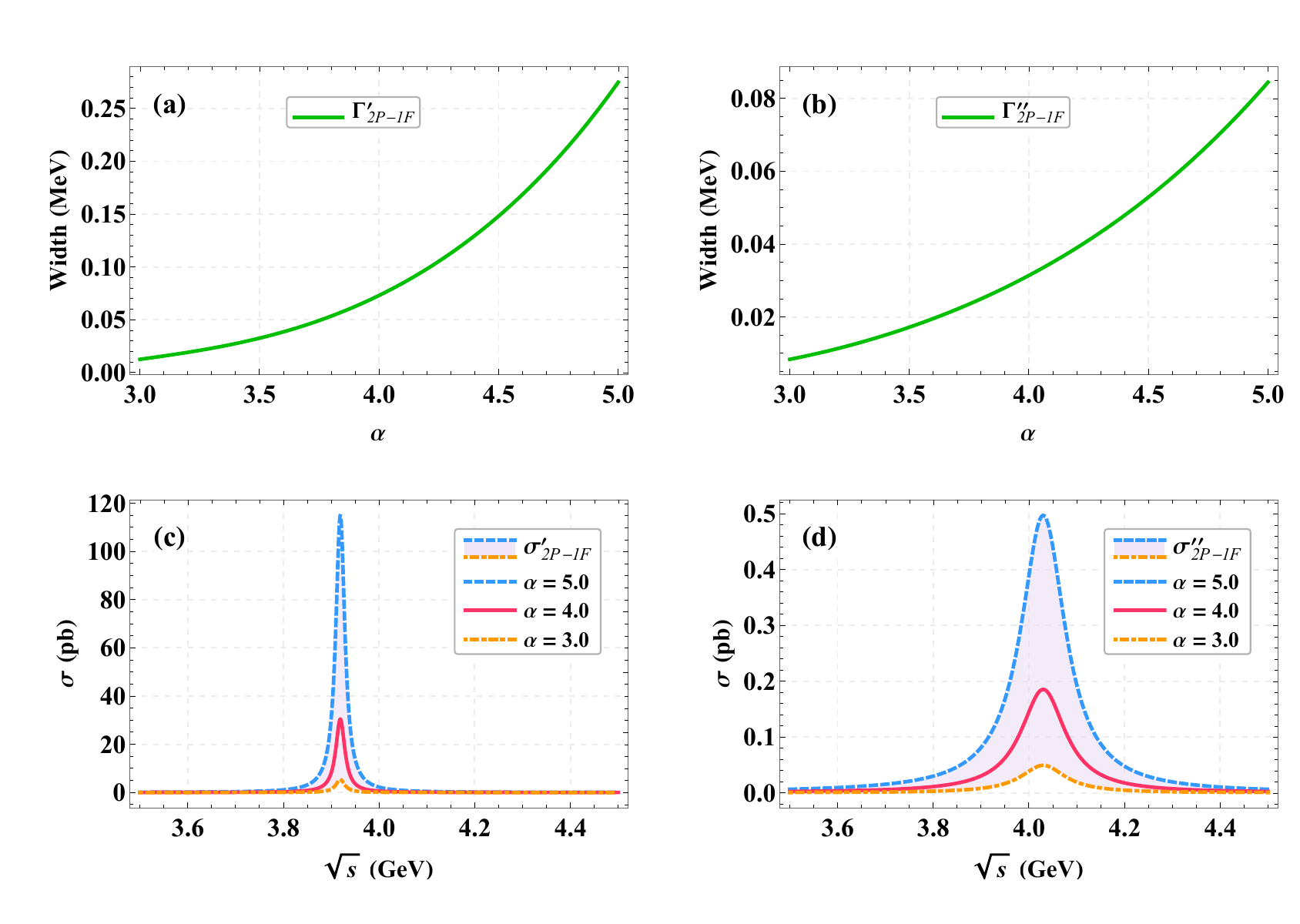}
	\end{tabular}
	\caption{
   Panels (a) and (b) show the decay widths of the low-mass and high-mass mixed states to $J/\psi \, \omega$ as functions of $\alpha$, respectively. Both widths exhibit an increasing trend with $\alpha$. Panels (c) and (d) display the total cross sections for the two-photon production of the low-mass and high-mass mixed states, respectively, followed by their decays to $J/\psi \, \omega$. The light purple band indicates the variation of the cross section as $\alpha$ varies from 3 to 5.}
	\label{Fig2}
\end{figure*}

In the actual calculation, we need to determine the parameter $\alpha$ in the form factor. Here, we adopt $\alpha$ in the range of 3 to 5\footnote{In Ref.~\cite{Chen:2013yxa}, the Lanzhou group found that the decay width $\Gamma(X(3915)\to\omega J/\psi)$ was consistent with experimental data when taking $\alpha = 4$. Given that the state we are calculating has a substantial 2$P$-wave component, we take $\alpha \in [3, 5]$.}, and the results are shown in Fig.~\ref{Fig2}.

The strong dependence on \(\alpha\) reflects the phenomenological nature of the hadronic-loop calculation. The form factor regularizes the loop integral, but
the present framework is not a fully renormalized field theory with
counterterms absorbing the cutoff dependence. The dipole form factor is a phenomenological
regularization scheme rather than a first-principles QCD result. Different
choices of form factors may lead to different absolute widths and cross
sections. The residual dependence on \(\alpha\) should therefore be regarded as
a model uncertainty of the hadronic-loop approach. Precise future
measurements may help
to constrain the cutoff parameter and make the predictions more quantitative.

In Fig.~\ref{Fig2} (a) and (b), we illustrate the numerical results of the decay widths $\Gamma^{\prime}_{2P-1F}$ and $\Gamma^{\prime\prime}_{2P-1F}$ as a function of the dimensionless parameter $\alpha$ in the hadronic loop mechanism. It is evident that both widths increase as $\alpha$ varies from 3.0 to 5.0, reflecting the sensitivity of the loop integral to the momentum cut-off scale. Specifically, the peak value of $\Gamma^{\prime}_{2P-1F}$ reaches approximately $0.27$ MeV, while $\Gamma^{\prime\prime}_{2P-1F}$ is suppressed to about $0.08$ MeV. Based on our theoretical framework, this discrepancy primarily stems from the input values of the coupling constants. As summarized in Table~\ref{tab:coupling_constants_updated}, the coupling constant $g_{\chi^{\prime}_{c2}DD^*}$ is $12.89$ GeV$^{-2}$, which is significantly larger than $g_{\chi^{\prime\prime}_{c2}DD^*} = 3.74$ GeV$^{-2}$.

Using the above results, we calculate the cross sections for the productions of the mixed $2P$-$1F$ charmonium states according to Eq.~(\ref{eq:21}) and present them in Fig.~\ref{Fig2} (c) and (d). The cross section $\sigma^{\prime}_{2P-1F}$ exceeds $100$ pb, while $\sigma^{\prime\prime}_{2P-1F}$ is suppressed to below $0.5$ pb, owing to the differences in both $\Gamma_{J/\psi\omega}$ and $\Gamma_{\gamma\gamma}$ between the two mixed states. 
Based on our predictions, we propose targeted strategies for searching these two mixed states via $\gamma\gamma \to J/\psi \omega$. The prominent $\chi^{\prime}_{c2}$ state should be readily observable at the Belle II experiment. Conversely, detecting the heavily suppressed $\chi^{\prime\prime}_{c2}$ state requires unprecedented statistics to overcome continuum backgrounds, making it an ideal target for the full Belle II dataset and the proposed Super Tau-Charm Facility (STCF).

It is meaningful to briefly discuss the cross section for $\gamma\gamma \to D\bar{D}$. Given that we have already calculated the $D\bar{D}$ widths for both mixed states using the QPC model, the cross sections can be readily obtained by applying Eq.~(\ref{eq:21}). According to our calculation, the cross section is 5.9~nb for the low-mass mixed state and 0.5~nb for the high-mass one. Both values suggest a promising potential for observation at the Belle~II experiment.


\section{Summary}
\label{sec:summary}

A key outcome of coupled-channel dynamics in heavy quarkonia is configuration mixing between states sharing the same $J^{PC}$ but differing in orbital angular momentum. While $S$-$D$ wave mixing, exemplified by the $\psi(3770)$ as a $2^3S_1$-$1^3D_1$ mixture \cite{Rosner:2001nm}, has been extensively studied and extended to higher excitations (e.g., in the $Y(4220)$ \cite{Wang:2019mhs,Man:2025zfu}, $Y(10753)$ \cite{Li:2021jjt}, and $\psi(4415)$ \cite{Wang:2019mhs}), mixing between $2P$- and $1F$-wave states has attracted far less attention, despite its potential relevance in the charmonium spectrum around 4.0 GeV. In the $J^{PC}=2^{++}$ sector, the $\chi_{c2}(2P)$ and the ground $F$-wave state $1^3F_2$ are predicted to lie close in mass, enabling significant mixing through intermediate hadronic loops even if the direct tensor force is weak.

In this work we go beyond the quenched approximation and investigate $2P$-$1F$ mixing driven by coupled-channel effects. Our analysis demonstrates that this dynamical mechanism induces sizable mixing angles, $\theta_1 = 7.5^\circ$ and $\theta_2 = 15.4^\circ$, indicating that the coupled-channel contribution dominates over the static tensor term. Using the resulting mixed-state wave functions, we compute the corresponding two-photon and two-gluon decay widths. Notably, the predicted $\Gamma_{\gamma\gamma}$ values differ significantly, with 0.14 keV for the lower-mass mixed state $\chi^{\prime}_{c2}$ and 0.05 keV for the higher-mass counterpart $\chi^{\prime\prime}_{c2}$, which provides a clear experimental observable for discriminating between the two configurations. Although current data are insufficient to extract the mixing angles precisely, our predictions provide concrete benchmarks for future high-precision measurements of the $\chi_{c2}(2P)$ and $\chi_{c2}(1F)$ charmonium states, thereby helping to clarify the fine structure of the charmonium spectrum in this energy domain.

We also focuses on the production of $2P$-$1F$ mixing states via $\gamma\gamma$ fusion. As shown in Fig.~\ref{Fig2}, the predicted widths and cross sections for the $J/\psi \, \omega$ final state of the two mixed states increase monotonically with $\alpha \in [3, 5]$, yet a striking magnitude gap exists between them. Specifically, at $\alpha=5$, the width $\Gamma^{\prime}_{2P-1F}$ reaches $0.27$ MeV, whereas $\Gamma^{\prime\prime}_{2P-1F}$ is only $0.08$ MeV. Correspondingly, the peak production cross section $\sigma^{\prime}_{2P-1F}$ exceeds $110$ pb, which is overwhelmingly larger than the heavily suppressed $\sigma^{\prime\prime}_{2P-1F}$ at merely $0.5$ pb. The predicted cross sections for the $D\bar{D}$ final state of the two mixed states $\chi^{\prime}_{c2}$ and $\chi^{\prime\prime}_{c2}$ are 5.9 nb and 0.5 nb respectively. We expect that future experiments will measure these cross sections precisely.



\begin{acknowledgments}
This work is also supported by the Natural Science Foundation of Gansu Province (No. 22JR5RA389, No. 25JRRA799), the National Natural Science Foundation of China under Grant Nos. 12335001 and 12247101, the ‘111 Center’ under Grant No. B20063, the fundamental Research Funds for the Central Universities, the project for top-notch innovative talents of Gansu province, and Lanzhou City High-Level Talent Funding.  T.L.G.
is supported by the Gansu Province Postgraduate Innovation Star Program No. 2026CXZX-042.
\end{acknowledgments}

\begin{appendix}

\onecolumngrid 

\section{Hadronic loop amplitudes} \label{Amplitudes}

Here, we list the amplitudes used in the hadronic loop calculation. For an amplitude, the subscript denotes the types of the two internal lines attached to the mixed state, and the superscript indicates the type of exchanged particle.
\begin{equation}
\scalebox{1}{$\displaystyle
\begin{aligned}
\mathcal{M}_{D\bar{D}}^{D} =& \int \frac{d^4q}{(2\pi)^4} \epsilon^{\rho \sigma}_{\chi^{\prime(\prime\prime)}_{c2}} \left[ {i g_{\chi^{\prime(\prime\prime)}_{c2}DD} (i q_{2\rho}) (i q_{1\sigma})} \right]  \frac{i}{q_1^2 - m_D^2} \frac{i}{q_2^2 - m_D^2} \frac{i}{q^2 - m_D^2} \\
&\times \left[ i g_{J/\psi DD} \epsilon_\psi^{*\eta} (-i q_{1\eta} - i q_\eta) \right] \left[ -i g_{DD\omega} \epsilon_\omega^{*\gamma}  (-i q_\gamma + i q_{2\gamma}) \right] \mathcal{F}^2(q^2),
\end{aligned}
$}
\label{eq:28}
\end{equation}

\begin{equation}
\scalebox{1}{$\displaystyle
\begin{aligned}
\mathcal{M}_{D\bar{D}}^{D^*} =& \int \frac{d^4q}{(2\pi)^4} \epsilon^{\rho \sigma}_{\chi^{\prime(\prime\prime)}_{c2}}  \left[ {i g_{\chi^{\prime(\prime\prime)}_{c2}DD} (i q_{2\rho}) (i q_{1\sigma})} \right] \frac{i}{q_1^2 - m_D^2} \frac{i}{q_2^2 - m_D^2} \frac{i \tilde{g}^{\delta\xi}(q)}{q^2 - m_{D^{*}}^2} \\
&\times \left[ -g_{J/\psi D^{*}D} \varepsilon_{\lambda\eta\tau\delta}  \epsilon^{*\eta}_{\psi}  (i p_\psi^\lambda)(i q^\tau) \right] \left[ -2 f_{D^{*}D\omega} \varepsilon_{\theta \gamma \phi \xi}  \epsilon^{*\gamma}  (i p_\omega^\theta)  (-i q^\phi + i q_2^\phi) \right]\mathcal{F}^2(q^2),
\end{aligned}
$}
\label{eq:29}
\end{equation}

\begin{equation}
\scalebox{1}{$\displaystyle
\begin{aligned}
\mathcal{M}_{D\bar{D}^{*}}^{D} =& \int \frac{d^4q}{(2\pi)^4} \epsilon^{\rho \sigma}_{\chi^{\prime(\prime\prime)}_{c2}}   \left[ {g_{\chi^{\prime(\prime\prime)}_{c2}DD^{*}} \varepsilon_{\rho\tau\lambda\theta} (-i p^\lambda) (i q_{2\sigma}) (i q_1^\theta)} \right]\frac{i}{q_1^2 - m_D^2} \frac{i \tilde{g}^{\tau\phi}}{q_2^2 - m_{D^{*}}^2} \frac{i}{q^2 - m_D^2} \\
&\times \left[ i g_{J/\psi DD} \varepsilon_\psi^{*\eta} (-i q_{1\eta} - i q_\eta) \right]  \left[ 2 f_{D^{*}D\omega} \varepsilon_{\delta \gamma \xi \phi} \epsilon_\omega^{*\gamma} (i p_\omega^\delta) (-i q^\xi + i q_2^\xi) \right]  \mathcal{F}^2(q^2),
\end{aligned}
$}
\label{eq:30}
\end{equation}

\begin{equation}
\scalebox{1}{$\displaystyle
\begin{aligned}
\mathcal{M}_{D\bar{D}^{*}}^{D^*} =& \int \frac{d^4q}{(2\pi)^4} \epsilon^{\rho \sigma}_{\chi^{\prime(\prime\prime)}_{c2}} \left[ { g_{\chi^{\prime(\prime\prime)}_{c2}DD^{*}} \varepsilon_{\rho\tau\lambda\theta} (-i p^\lambda) (i q_{2\sigma}) (i q_1^\theta)} \right]  \frac{i}{q_1^2 - m_D^2} \frac{i \tilde{g}^{\tau\phi}}{q_2^2 - m_{D^{*}}^2} \frac{i \tilde{g}^{\delta\xi}}{q^2 - m_{D^{*}}^2}  \left[ -g_{J/\psi D^{*}D} \varepsilon_{\omega \eta \chi \delta} \epsilon_\psi^{*\eta} (i p_\psi^\omega)  (i q^\chi) \right]  \\
&\times \Bigl[ i g_{D^{*}D^{*}\omega} \epsilon_\omega^{*\gamma}(-i q_\gamma + i q_{2\gamma}) g_{\phi\xi}  + 4 i f_{D^{*}D^{*}\omega} \epsilon_\omega^{*\gamma} (i p_\omega^\mu) g_{\mu\phi} g_{\gamma\xi}  - 4 i f_{D^{*}D^{*}\omega} \epsilon_\omega^{*\gamma} (i p_\omega^\nu) g_{\nu\xi} g_{\gamma\phi} \Bigl] \mathcal{F}^2(q^2),
\end{aligned}
$}
\label{eq:31}
\end{equation}

\begin{equation}
\scalebox{1}{$\displaystyle
\begin{aligned}
\mathcal{M}_{D^*\bar{D}}^{D} =& \int \frac{d^4q}{(2\pi)^4} \epsilon^{\rho \sigma}_{\chi^{\prime(\prime\prime)}_{c2}} \left[ { -g_{\chi^{\prime(\prime\prime)}_{c2}DD^*} \varepsilon_{\rho\lambda\tau\phi} (-i p^\tau) (i q_2^\phi) (i q_{1\sigma})} \right]  \frac{i \tilde{g}^{\lambda\theta}}{q_1^2 - m_{D^{*}}^2} \frac{i}{q_2^2 - m_D^2} \frac{i}{q^2 - m_D^2} \\
&\times \left[ -g_{J/\psi D^{*}D} \varepsilon_{\delta \eta \xi \theta} \epsilon_\psi^{*\eta}(i p_\psi^\delta) (-i q_1^\xi) \right]  \left[ -i g_{DD\omega} \epsilon_\omega^{*\gamma} (-i q_\gamma + i q_{2\gamma}) \right]  \mathcal{F}^2(q^2),
\end{aligned}
$}
\label{eq:32}
\end{equation}

\begin{equation}
\scalebox{1}{$\displaystyle
\begin{aligned}
\mathcal{M}_{D^*\bar{D}}^{D^*} =& \int \frac{d^4q}{(2\pi)^4} \epsilon^{\rho \sigma}_{\chi^{\prime(\prime\prime)}_{c2}} \left[ {-g_{\chi^{\prime(\prime\prime)}_{c2}DD^*} \varepsilon_{\rho\lambda\tau\phi} (-i p^\tau) (i q_2^\phi) (i q_{1\sigma})} \right]  \frac{i \tilde{g}^{\lambda\theta}}{q_1^2 - m_{D^{*}}^2} \frac{i}{q_2^2 - m_D^2} \frac{i \tilde{g}^{\delta\xi}}{q^2 - m_{D^{*}}^2} \left[ -2 f_{D^{*}D\omega} \varepsilon_{m \gamma a b} \epsilon_\omega^{*\gamma} (i p_\omega^m) (-i q^a + i q_2^a) g_{b\xi} \right]  \\
&\times \Bigl\{ -i g_{J/\psi D^{*}D^{*}} \Bigl[ \epsilon_\psi^{*\eta}(-i q_{1\eta} - i q_\eta) g_{\delta\theta} + \epsilon_\psi^{*\eta} (i p_\psi^\chi + i q_1^\chi) g_{\eta\theta} g_{\chi\delta} + \epsilon_\psi^{*\eta} (i q^\chi - i p_\psi^\chi) g_{\eta\delta} g_{\chi\theta} \Bigl] \Bigl\} \mathcal{F}^2(q^2),
\end{aligned}
$}
\label{eq:33}
\end{equation}
where the subscript of the amplitude denotes the two internal lines connected to the $\chi^{\prime(\prime\prime)}_{c2}$, while the superscript represents the exchanged particle.

The total amplitude for $\chi^{\prime(\prime\prime)}_{c2}$ decaying to the $J/\psi \, \omega$ final state is written as:
\begin{equation}
\begin{aligned}
\mathcal{M} [\chi^{\prime(\prime\prime)}_{c2} \to J/\psi \omega] = & 4\; \Bigl[ \mathcal{M}_{D\bar{D}}^{D} + \mathcal{M}_{D\bar{D}}^{D^*} + \mathcal{M}_{D\bar{D}^{*}}^{D} + \mathcal{M}_{D\bar{D}^{*}}^{D^*} + \mathcal{M}_{D^*\bar{D}}^{D} + \mathcal{M}_{D^*\bar{D}}^{D^*} \Bigl],
\end{aligned}
\label{eq:34}
\end{equation}
where the factor 4 results from the charge conjugate and isospin transformations. 

\twocolumngrid

\end{appendix}

\bibliography{ref.bib}

\end{document}